\documentstyle[12pt,rotate,cite,epsfig]{article}
\newcommand{\beq}{\begin{equation}}
\newcommand{\eeq}{\end{equation}}
\newcommand{\bea}{\begin{eqnarray}}
\newcommand{\eea}{\end{eqnarray}}
\newcommand{\bd}{\begin{displaymath}}
\newcommand{\ed}{\end{displaymath}}

\newcommand{\ete}{\eta^{\prime} }

\begin{document}
\bibliographystyle{physics}
\renewcommand{\thefootnote}{\fnsymbol{footnote}}

\author{
Dongsheng Du${}^{1,2}$~~~ Yadong Yang${}^{1,2,3}$~~~ Guohuai Zhu${}^{2}$
\thanks{Email: duds@bepc3.ihep.ac.cn, yangyd@hptc5.ihep.ac.cn,zhugh@hptc5.
ihep.ac.cn}\\
{\small\sl ${}^{1}$ CCAST (World Laboratory), P.O.Box 8730, Beijing
100080, China}\\
{\small\sl ${}^{2}$ Institute of High Energy Physics, Academia Sinica,
 P.O.Box 918(4), Beijing 100039, China\thanks{Mailing address} }\\
{\small\sl ${}^{3}$ Physics Department of Henan Normal University, Xingxiang,
Henan 453002, China}\\
}
\date{}
\title{
{\large\sf
\rightline{BIHEP-Th/98-5}
}
\vspace{1cm}
{\LARGE\sf Further Analysis of $B \rightarrow \ete K(K^*),\eta
K(K^*)$ 
Processes\thanks{Supported in part by National Natural Science
Foundation of China} }}

\maketitle
\thispagestyle{empty}
\begin{abstract}
\noindent
Combining three mechanisms, we reanalysis processes of 
$B \rightarrow \ete K(K^*),\eta K(K^*)$ and calculate their branching
ratios. The results are 
compared with other mechanisms in the literature. The striking feature of
the gluon fusion mechanism is emphasized and its experimental test is
discussed.
\end{abstract}

\newpage
\setcounter{page}{1}

\setcounter{footnote}{0}
\renewcommand{\thefootnote}{\arabic{footnote}}
\section{Introduction}

Since the CLEO collaboration reported large  branching fractions
\cite{cleo1}:
\bea
{\cal BR}(B^{\pm} \rightarrow  \eta^{\prime} X_s)
&=&(6.2 \pm 1.5 \pm 1.3)\times 10^{-4}  (2.0 < P_{\ete} < 2.7GeV),\\
{\cal BR}(B^{\pm} \rightarrow  \eta^{\prime} K^{\pm})
&=&(6.5^{+1.5}_{-1.4}\pm 0.9)\times 10^{-5}~.
\eea
It has received much attention because it is a great experimental 
achievement in rare B decays which are dominated by penguin contributions
and the branching fractions are surprisingly large compared with earlier
theoretical estimation.

Now it seems that we have a qualitatively reasonable interpretation of the
semi-inclusive
decay $B \rightarrow \ete X_s $: It is due to the special property of $\ete$
which has anomalously large coupling to gluons. But as to the exclusive decay
$B \rightarrow \ete K$ there are still several possible mechanisms. This is
partly because it is less reliable to estimate exclusive decay
than to semi-inclusive decay due to our complete ignorance of the hadronization
process.

In the
following we first carefully estimate the contribution from the conventional
mechanism. We find that the terms, such as
$\langle ~ \ete \mid \bar s \gamma_5 s \mid 0\rangle ~\langle ~K^- \mid \bar s b \mid B^- \rangle $
, are dominant terms if using Dirac equation to estimate them,
 but the large factors such as
$\frac{m_{\ete}}{m_s}$ have large uncertainties because the quark mass
is uncertain, so we also use PQCD method to estimate them and get 
comparatively small result which indicates that we need new mechanisms
to account for the experiment result.
Secondly we discuss some possible new mechanisms.
We find that if
$b \rightarrow \bar c cs \rightarrow s \ete $ is the dominant mechanism,
the branching fraction of $B \rightarrow \ete K^*$ would be too large to
fit the CLEO data.
 Thirdly we try to combine the
mechanisms proposed in ref[6](two gluons fusion) , ref[7](non-zero
$f^c_{\ete}$ contribution) and the conventional mechanism to compute
the branching fractions of
$B \rightarrow \ete K(K^*),\eta K(K^*)$. The results are
compared with other mechanisms in the literature.

The article is organized as follows: Section 2 gives the theoretical formula
for the decay amplitude of
$B \rightarrow \ete K$ using factorization assumption and Dirac equation, and
some uncertainties in the formula are briefly discussed.
Section 3 is devoted to the PQCD method to estimate the hadronic
matrix elements such as 
$\langle ~ \ete \mid \bar s \gamma_5 s \mid 0\rangle ~\langle ~K^- \mid \bar s b \mid B^- \rangle $
. In section 4, some possible mechanisms are discussed and used
to fit the data. Section 5 is for the concluding remarks.

\section{Preview}
 The standard theoretical frame to estimate the non-leptonic B decays
  is based on the effective Hamiltonian
\bea
H_{eff}&=&\frac{G_F}{\sqrt{2}}
\left[
\lambda_u (C_1 O^u_1 +C_2 O^u_2 )+\lambda_c (C_1 O^c_1 +C_2 O^c_2 )
-\lambda_t \sum_{i=3}^{10}C_i O_i 
\right]
\eea
( $\lambda_i =V_{ib}V_{is}^*$) and BSW model \cite{bsw}. But there exists large
uncertainties in estimating the hadronic matrix elements like
\bea
\langle ~\ete \mid \bar s \gamma_5 s \mid 0\rangle ~,~~~\langle ~K^- \mid \bar s b \mid B^- \rangle ~.
\eea
In the literature Dirac equation is used to estimate such matrix elements:
\bea
\langle ~\ete \mid \bar s \gamma_5 s \mid 0\rangle ~&=&i\frac{ m^2_{\ete}}{2m_s}f_{\ete}^s
~,\\
\langle ~K^- \mid \bar s b \mid B^- \rangle ~&=&\frac{ P_{\ete}^{\mu}}{m_b -m_s}
\langle ~K^- \mid \bar s \gamma_{\mu}b \mid B^- \rangle ~~.
\eea
With these relations one can write down the amplitude of $B^- \rightarrow\ete K^- $ \cite{chenghy1}:
\bea
{\cal A}(B^- \rightarrow \ete K^- )=\frac{G_F}{\sqrt{2}}
\{
V_{ub}V_{cs}^* (a_1^{eff}X_1+a_2^{eff}X_{2u}+a_1^{eff}X_3)
+V_{cb}V_{cs}^* a_2^{eff}X_{2c}\nonumber\\
-V_{tb}V_{ts}^*
\left[
(a_4 +a_{10}+2(a_6 +a_8 )\frac{ m_K^2}{(m_s +m_u )(m_b -m_u )})X_1
\right.\nonumber\\
\left.
+(2a_3 -2a_5 -\frac{1}{2}a_7 +\frac{1}{2}a_9 )X_{2u}
+(a_3 -a_5 -a_7 +a_9 )X_{2c}\right.\nonumber\\
\left.
+(a_3 +a_4 -a_5 +\frac{1}{2}a_7 -\frac{1}{2}a_9 -\frac{1}{2}a_{10}
  +(a_6 -\frac{1}{2}a_8 )\frac{ m_{\ete}^2}{m_s (m_b -m_s )})X_{2s}
\right.\nonumber\\
\left.
+(a_4 +a_{10}+2(a_6 +a_8 )\frac{m^2_B}{(m_s -m_u )(m_b +m_u )})X_3
\right]
\}~.
\eea
where $X_i$ are factorizable terms
\bea
X_1&=\langle ~K^- \mid (\bar s u)_{V-A} \mid 0\rangle ~\langle ~\ete \mid (\bar u b)_{V-A} \mid B^- \rangle ~
,\nonumber\\
X_{2q}&=\langle ~\ete \mid (\bar q q)_{V-A} \mid 0\rangle ~\langle ~K^- \mid (\bar s b)_{V-A} \mid
B^- \rangle ~,\\
X_3&=\langle ~\ete K^- \mid (\bar s u)_{V-A} \mid 0\rangle ~\langle ~0 \mid (\bar u b)_{V-A}\mid B^-\rangle ~
.\nonumber
\eea
and
\bea
\langle ~K^- \mid \bar s \gamma_5 u \mid 0\rangle ~\langle ~\ete \mid \bar u b \mid B^- \rangle ~
=-\frac{ m_K^2}{(m_s +m_u )(m_b -m_u )}X_1, \nonumber \\
\langle ~\ete \mid \bar s \gamma_5 s \mid 0\rangle ~\langle ~K^- \mid \bar s b \mid B^- \rangle ~
=-\frac{ m_{\ete}^2}{2m_s (m_b -m_s )}X_{2s},  \\
\langle ~\ete K^- \mid \bar s u \mid 0\rangle ~\langle ~0 \mid \bar u \gamma_5 b \mid B^-\rangle ~
=-\frac{m^2_B}{(m_s -m_u )(m_b +m_u )}X_3.\nonumber
\eea
But the large factors such as $\frac{m_{\ete}}{m_s}$ in the amplitude
have large uncertainties because the quark mass is uncertain.

Through the numerical calculation, we find  Equation(7) gives a branching
ratio of
$3.5 \times 10^{-5}$,
and if we neglect the term
$\frac{m^2_{\ete}}{m_s (m_b -m_s )}X_{2s}$ in the equation(7), the result is
$6.63\times 10^{-6}$, which is too small. So we find that such a term is very
important, and we try to find an alternative way to estimate such matrix 
elements.\\

\section{Perturbative QCD calculation}
Brodsky {\it et al. }\cite{brod} has pointed out that because of the large momentum
transfers involved in the decays of B to light mesons, 
the factorization formula of PQCD for exclusive reactions
 becomes applicable: the amplitude can be written as a convolution of a
 hard-scattering quark-gluon amplitude $T_h$, and meson distribution amplitudes
$\phi (x,Q^2 )$ which describe the fractional longitudinal momentum
distribution of the quark and antiquark in each meson. An important feature
of this formalism is that, at high momentum tranfer, long-range final state
interactons between the outgoing hadrons can be neglected. In the case of
nonleptonic weak decays the mass of the heavy meson $M_H^2 $ establishes
the relevant momentum scale $Q^2 \sim M_H^2 $, so that for a sufficiently
massive initial state the decay amplitude is of order $\alpha_s (Q^2 )$, even
without including loop corrections to the weak hamiltonian. The dominant 
contribution is controlled by single gluon exchange.

We try to use PQCD method to estimate those hadronic matrix elements
such as $\langle ~\ete \mid \bar s \gamma_5 s \mid 0\rangle ~$ and
$\langle ~K^- \mid \bar s b \mid B^- \rangle ~$.
The wave functions of $B^- $ and $K^- $ are:
\bea
\Psi_B(x) &=\frac{1}{\sqrt{2}} \frac{I_C}{\sqrt{3}} \phi_B (x)
(\slash{\hskip -3mm}P_B +M_B )\gamma_5 ~, \nonumber\\
\Psi_K(y) &=\frac{1}{\sqrt{2}} \frac{I_C}{\sqrt{3}} \phi_K (y)
(\slash{\hskip -3mm}P_K +M_K )\gamma_5 ~,
\eea
where $I_C$ is an identity in color space. In QCD, the integration of the
distribution amplitude is related to the meson decay constant 
\bea
\int \phi_K (y)dy=\frac{1}{2\sqrt{6}}f_K ~,~~~~
\int \phi_B (x)dx=\frac{1}{2\sqrt{6}}f_B ~.
\eea
Then we can write down the amplitude of Fig.1 as
\begin{eqnarray}
\langle ~K^- \mid \bar s \gamma_{\mu}\gamma_5 u\mid 0\rangle ~_{PQCD} 
&= 3\times \frac{1}{\sqrt{2}}\frac{1}{\sqrt{3}}
\int dy\phi_K (y) Tr \left[\gamma_5~(\slash{\hskip -3mm}q_K +M_K )
\gamma_{\mu}\gamma_5 \right]& =f_K q_{\mu}~,\nonumber\\
\langle ~K^- \mid \bar s \gamma_5 u \mid 0\rangle ~_{PQCD}&=
3\times \frac{1}{\sqrt{2}}\frac{1}{\sqrt{3}}
\int dy\phi_K (y) Tr \left[\gamma_5~(\slash{\hskip -3mm}q_K +M_K )
\gamma_5 \right]&=f_K M_K~.
\end{eqnarray}
Compared with the widely used results
\begin{eqnarray}
\langle ~K^- \mid \bar s \gamma_{\mu}\gamma_5 u\mid 0\rangle ~&=-i~f_k
q_{\mu},\nonumber\\
\langle ~K^- \mid \bar s \gamma_5 u \mid 0\rangle ~_{Dirac}&=i\frac{
M^2_K}{(m_s +m_u )}f_K,
\end{eqnarray}
so, except for a $-i$ factor,
$\langle ~K^- \mid \bar s \gamma_{\mu}\gamma_5 u\mid 0\rangle ~_{PQCD}=
\langle ~K^- \mid \bar s \gamma_{\mu}\gamma_5 u\mid 0\rangle ~,$
but $\langle ~K^- \mid \bar s \gamma_5 u \mid 0\rangle ~_{PQCD}~\ll~\langle ~K^- \mid \bar s \gamma_5 u
\mid 0\rangle ~_{Dirac}$
. Similarly, we can get
\bea
\langle ~\ete \mid \bar q \gamma_{\mu}\gamma_5 q\mid 0\rangle ~_{PQCD}
=f_{\ete}^q p_{\mu},~~~~~~
\langle ~\ete \mid \bar q \gamma_5 q\mid 0\rangle ~_{PQCD}
=f_{\ete}^q M_{\ete}.
\eea
In a consistent way, we can use perturbative QCD to estimate the matrix
elements like $\langle ~K^- \mid \bar s \gamma_{\mu}b \mid B^- \rangle ~$ and
$\langle ~K^- \mid \bar s b \mid B^- \rangle ~$(Fig.2, Fig.3),
we have neglected the fermi motion of quarks, while the gluons in the
Fig.2,3 are hard because
\bea
k^2 =(xP_B -(1-y)P_K )^2 \simeq -x(1-y)M_B^2 \sim 1 GeV^2
\eea
so, we can use perturbative QCD method to calculate the amplitude and it
turns out to be
\begin{eqnarray}
&\langle &K^- \mid \bar s \gamma_{\mu}b \mid B^- \rangle ~_{PQCD}= -\frac{2}{3}g^2
\int dxdy \phi_B (x)\phi_K (y) \\
&\{&
\frac{Tr \left[\gamma_5 (\slash{\hskip -3mm}q+M_K )\gamma^{\nu}
\slash{\hskip -3mm}P_s \gamma_{\mu} (\slash{\hskip -3mm}P +M_B)
\gamma_5 \gamma_{\nu} \right]}{k^2 P_s^2 }
+\frac{Tr \left[\gamma_5 (\slash{\hskip -3mm}q+M_K )\gamma_{\mu}
(\slash{\hskip -3mm}P_b +m_b )\gamma^{\nu}(\slash{\hskip -3mm}P +M_B)
\gamma_5 \gamma_{\nu} \right]}{k^2 (P_b^2 -m_b^2 )}
\}~.\nonumber
\end{eqnarray}
In order to get the quantitative estimation, we take the wave functions as
\cite{brod,brodsky1} 
\bea
\phi_B (x)=\frac{f_B}{2\sqrt{6}}\delta(x-\epsilon_B )~,~~~~~~~~~~~
\phi_K (y)=\sqrt{\frac{3}{2}}f_K y(1-y) ~.
\eea
 we get
\bea
\langle ~K^- \mid \bar s \gamma_{\mu}b \mid &B^-& \rangle ~_{PQCD}=
\frac{4\pi}{3}\alpha_s f_B f_K \left( \frac{2M_B M_K }{\epsilon_B^2 M_B^4}+
\right. \nonumber\\
& &\left. \int dy y\frac{-4yM_B^2 +8yM_B M_K +8m_b M_B -4M_B m_K}
{-\epsilon_B^2 M_B^2 (yM_B^2-m_b^2 )}\right)
(p+q)_{\mu}\nonumber\\
& &= F (p+q)_{\mu},
\eea
where we have neglected another term $ F^{'}(p-q)_{\mu}$ which has negligible
contribution to the amplitude
$\langle ~\ete \mid \bar s \gamma^{\mu}\gamma_5 s\mid 0\rangle \langle ~K^- \mid \bar s \gamma_{\mu}b \mid B^- \rangle.$
For simplicity, we will drop such terms like $F^{'}(p-q)_{\mu}$
in the following, since their contributions always proportion to
$m_{\ete}^2,~m_k^2$ etc. The numerical results for F are presented in Table 1,
where we have taken
\bea
\alpha_s(\langle k^2\rangle ) \simeq 0.38, f_B =200MeV, f_K =160MeV.\nonumber
\eea
From Table 1, we find that our PQCD results are
sensitive to the values of parameter $\epsilon_B$ and $m_b$, and seem small 
compared with the BSW result:
\bea
\langle ~K^- \mid \bar s \gamma_{\mu}b \mid B^- \rangle ~&=F_0^{BK}(m^2_{\ete})(p+q)_{\mu}
= 0.35(p+q)_{\mu}.
\eea
In fact the PQCD results are comparatively small in many cases, however,
we believe that the ratio
\bea
Rm=\frac{\langle ~\ete \mid \bar s  \gamma_5 s \mid 0\rangle ~_{PQCD}\langle ~K^- \mid \bar s b \mid
B^- \rangle ~_{PQCD}}{\langle ~\ete \mid \bar s \gamma_{\mu}\gamma_5 s \mid 0\rangle ~_{PQCD}
\langle ~K^- \mid \bar s \gamma^{\mu}b \mid B^- \rangle ~_{PQCD}}
\eea
is more reliable because
main uncertainties can be canceled. We can write down the matrix element
$\langle ~K^- \mid \bar s b \mid B^- \rangle $ as
\bea
\langle ~K^- \mid \bar s b \mid B^- \rangle ~_{PQCD}&=&-\frac{2}{3}g^2
\int dxdy \phi_B (x)\phi_K (y)
\{
\frac{Tr \left[\gamma_5 (\slash{\hskip -3mm}q+M_K )\gamma^{\nu}
\slash{\hskip -3mm}P_s (\slash{\hskip -3mm}P +M_B)
\gamma_5 \gamma_{\nu} \right]}{k^2 P_s^2 }\nonumber\\
& &+\frac{Tr \left[\gamma_5 (\slash{\hskip -3mm}q+M_K )
(\slash{\hskip -3mm}P_b +m_b )\gamma^{\nu}(\slash{\hskip -3mm}P +M_B)
\gamma_5 \gamma_{\nu} \right]}{k^2 (P_b^2 -m_b^2 )}
\} \nonumber\\
&=&\frac{8\pi}{3}\alpha_s f_B f_K 
\{
\frac{M_K M_B^2 (1-2\epsilon_B )- M_B (M_K^2-\frac{x}{2}M_B^2 )}
{\epsilon_B^2 M_B^4 }\nonumber\\
&~&+ \int dy y\frac{4m_b M_B M_K -m_b M_B^2 
-(1+y)M_K M_B^2 +8M_B^3 }{-\epsilon_B M_B^2 (yM_B^2-m_b^2 )}
\}.
\eea 
With eq.(14), (18), (21),
we find that the ratio is really not sensitive to the values of parameter
$\epsilon_B $ and $m_b $.
\bea
Rm=(0.19\sim 0.20)
(0.05\leq \epsilon_B \leq 0.08,~~4.8GeV\leq m_b \leq 5.0GeV)
\eea
If our results are right, the ratio where Dirac equation is used,
\bea
\frac{\langle ~\ete \mid \bar s  \gamma_5 s \mid 0\rangle ~_{Dirac}\langle ~K^- \mid \bar s b \mid
B^- \rangle ~_{Dirac}}{\langle ~\ete \mid \bar s \gamma_{\mu}\gamma_5 s \mid 0\rangle ~
\langle ~K^- \mid \bar s \gamma^{\mu}b \mid B^- \rangle ~}
=\frac{m_{\ete}^2}{2m_s (m_b -m_s )}
\simeq  0.9\\
(m_s (m_b )\simeq 105mev,~~4.8GeV\leq m_b \leq 5.0GeV)\nonumber
\eea
may be misleadingly large.

It is difficult to quantitatively estimate the hadronic matrix element
$\langle ~\ete \mid \bar u b \mid B^- \rangle ~$ because we don't know the wave function of
$\ete$. Fortunately in the calculation of $\langle ~K^- \mid \bar s b \mid B^- \rangle ~$
we find that $\langle ~K^- \mid \bar s b \mid B^- \rangle ~_{PQCD} \sim
\langle ~K^- \mid \bar s b \mid B^- \rangle ~_{Dirac}$
, so we can reasonably assume that 
$\langle ~\ete \mid \bar u b \mid B^- \rangle ~_{PQCD} \sim \langle ~\ete \mid \bar u b \mid B^- \rangle ~_{Dirac}$
and get the estimation:
\bea
\frac{\langle ~K^- \mid \bar s \gamma_5 u \mid 0\rangle ~_{PQCD}
\langle ~\ete \mid \bar u b \mid B^-\rangle ~_{PQCD}}
{\langle ~K^- \mid \bar s \gamma^{\mu}\gamma_5 u \mid 0\rangle ~_{PQCD}
\langle ~\ete \mid \bar u \gamma_{\mu}b \mid B^-\rangle ~_{PQCD}}
\sim 0.1 .
\eea
Because the term 
 \bea
 \langle ~K^- \mid \bar s \gamma_5 u \mid 0\rangle ~
\langle ~\ete \mid \bar u b \mid B^-\rangle ~ =
\frac{-M_K^2 }{(m_s +m_u )(m_b -m_u )}X_1
\eea
in eq.(7)(using Dirac equation) 
only contributes about 10 \% to the total amplitude,
our PQCD estimation (eq.24) will not substantially change
the total amplitude though it is not very precise.

For the hadronic matrix element 
\bea
\langle ~\ete K^- \mid \bar s u \mid 0\rangle ~\langle ~0 \mid \bar u \gamma_5 b \mid B^- \rangle ~
=-\frac{m_B^2}{(m_s-m_u)(m_b+m_u)}X_3~~(using~Dirac~equation) 
\eea
It contribute about 10 \% to the total amplitude in eq.(7) although it
has an uncomfortable large factor $\frac{m_B}{(m_s-m_u)}$, so we can 
still neglect it safely.

From the above analysis, we obtain the revised conventional amplitude
of $B^- \rightarrow \ete K^- $ (assume $f_{\ete}^c=0$) 
\bea
{\cal M}_{r.c}(B^- \rightarrow \ete K^- )&=&\frac{G_F}{\sqrt{2}}
\{
V_{ub}V_{cs}^* (a_1^{eff}X_1+a_2^{eff}X_{2u})
\nonumber\\
& &-V_{tb}V_{ts}^*
\left[
(a_4 +a_{10}+0.2(a_6 +a_8 )
)X_1\right. \nonumber\\
& & +(2a_3 -2a_5 -\frac{1}{2}a_7 +\frac{1}{2}a_9 )X_{2u}
\nonumber\\
& &\left. +(a_3 +a_4 -a_5 +\frac{1}{2}a_7 -\frac{1}{2}a_9 -\frac{1}{2}a_{10}
  +0.20(2a_6 -a_8 ))X_{2s}
\right]
\}~.
\eea
Taking $V_{tb}V_{ts}^* =-0.039 $, we get 
\bea
{\cal BR}(B^- \rightarrow \ete K^- ) \simeq 0.9 \times 10^{-5}
\eea
(which is much smaller than the result of $3.5 \times 10^{-5}$ 
where Dirac equation is employed.)
So we find that the conventional decay mechanism can not account for the 
experimental data. We need some new mechanisms to interpret the experiment.

\section{Some new mechanisms}
There are several mechanisms which may enhance the decay rate of
 $B^{\pm} \rightarrow \ete K^{\pm} $, they all use the unique properties of
$\ete$\cite{chenghy1}:\\
\begin{description}
\item[(i)] $b \rightarrow sg^* $ and $g^* \rightarrow g \ete $ 
\cite{soni} via QCD anomaly , it is important to the inclusive decay
$B \rightarrow \ete X_s $, but it seems difficult to realize its contribution
to two$-$body exclusive decay $B^{\pm} \rightarrow \ete K^{\pm} $.
\item[(ii)] $b \rightarrow sgg \rightarrow s \ete $\cite{ali}. Because the
effective vertex of
 $b \rightarrow sgg $ is very small, it is impossible to account for the
 large $B^{\pm} \rightarrow \ete K^{\pm} $ branching ratio.
\item[(iii)] Zhitnitsky \cite{zhit} have proposed a hopeful mechanism that
$\ete$ can be directly produced through 
$b \rightarrow \bar c cs \rightarrow \ete s$ if
$\langle ~\ete \mid \bar c \gamma^{\mu}\gamma_5 c \mid 0\rangle ~=-i f_{\ete}^c
P_{\ete}^{\mu}\neq 0$
. Unfortunately we can't determine
$f_{\ete}^c $ precisely from QCD. If this mechanism dominates over
other mechanisms of the decay $B^{\pm} \rightarrow \ete K^{\pm} $ ,
Zhitnitsky gets
$f_{\ete}^c=140 MeV$. However one would find
\bea
 \frac{ {\cal BR}(B^- \rightarrow \ete K^{*-} )}
     { {\cal BR}(B^- \rightarrow \ete K^ - )}
=\left|\frac{X_{2c}^{\prime}}{X_{2c}}\right |^2 
=0.83.
\eea
and the branching ratio for
$B^{\pm} \rightarrow \ete K^{*\pm} $ would exceed the upper limit of the
experimental data. Furthermore, if we take the contribution from the
conventional mechanism into account we find that the branching ratio
${\cal BR}(B^{\pm} \rightarrow \ete K^{*\pm}) $ is larger than
${\cal BR}(B^{\pm} \rightarrow \ete K^{\pm}) $, this strongly disagrees with
the experiment \cite{cleo1}.
So it is unlikely to be the dominant decay mechanism. In the following
we tend to treat $f_{\ete}^c $ as a free parameter in the range of
$-$40 MeV to 40 MeV to take into account the contributions of this mechanism.
\item[(iv)] Another possible mechanism is proposed in \cite{yangyd}.
This mechanism is motivated by the fact that both the recoil between $\ete$
and $K^-$ and the energy released in the process are large. The gluon from
$b \rightarrow sg $ vertex would carry energy about $M_B/2$ and then
materializes to $\ete$ and emits another hard gluon to balance color and
momentum. This is a hard scattering process, and the perturbative
 QCD method is applicable. This mechanism is depicted in Fig.4 .
 We re-examine its contributions and the numerical results are presented
in Table 2. And we find that this mechanism is very important.

\end{description}
 
We do not discuss some other new mechanisms which involve new physics beyong 
the SM. We think that the contributions of the SM should be carefully examined first.
So, in what follows, we try to combine the mechanisms (iii) and (iv)  with 
the revised conventional amplitude eq.(28). 
The amplitude of mechanism (iv) is \cite{yangyd}
\bea
{\cal M}_{iv}=\frac{G_F}{\sqrt{2}}\alpha_s^2 C_{eff}C_F 32 (V_{is}V_{ib}^*)
~~~~~~~~~~~~~~~~~~~~~~~~~~~~~~~~~~~~~~~~~~~~~~~~~~~~~~~~~~~~~~\nonumber\\
\times\int dxdy\phi_{B}(x)\phi_{K}(y)
\frac{
F_1^i k_1^2 [p\cdot k_1 q\cdot k_2 -p\cdot k_2 q\cdot k_1 ]
+F_2^i M_B m_b [q\cdot k_2 k_1^2 -q\cdot k_1 k_1 \cdot k_2 ] }
{k_{1}\cdot k_2  k_1^2 k_2^2 },
\eea
and the amplitude of mechanism (iii) is
\bea
{\cal M}_{iii}=\frac{G_F}{\sqrt{2}}
\left[
V_{cb}V_{cs}^* a_2^{eff}X_{2c}-V_{tb}V_{ts}^*
(a_3 -a_5 -a_7 +a_9 )X_{2c}
\right]
\eea
The total amplitude is 
\bea
{\cal M}_{total}={\cal M}_{r.c}+{\cal M}_{iii}+{\cal M}_{iv}
\eea
The numerical results are listed in Table 2, where we have taken
$f_8=1.38f_{\pi},~~f_0=1.06 f_{\pi}. $ It is shown that 
a nonzero $f_{\ete}^c$ may be important to explain the experimental
data. From Table 2, one may note that $f_{\ete}^c=-30 MeV$ seems fit
the experiment better than $f_{\ete}^c=-15 MeV$. However, considering
the uncertainties in estimating the branching
ratios, for example the PQCD method often gives comparatively small result,
we choose $f_{\ete}^c=-15 MeV$ as an input parameter to 
estimate some other related channels, such as
\bea
B^- \rightarrow \ete K^{*-},~~~~
B^- \rightarrow \eta K^-,~~~~
B^- \rightarrow \eta K^{*-},\nonumber
\eea
\bea
B^0 \rightarrow \ete K^{*0},~~~~
B^0 \rightarrow \eta K^0,~~~~
B^0 \rightarrow \eta K^{*0}.\nonumber
\eea
 We present a complete expression of the amplitude of
$B^{\pm} \rightarrow \ete K^{*\pm}$ as an example:
\bea
{\cal M}_{total}^{'}={\cal M}_{r.c}^{'}+{\cal M}_{iii}^{'}
+{\cal M}_{iv}^{'},
\eea
where
\bea
{\cal M}_{r.c}^{'}&=&\frac{G_F}{\sqrt{2}}
\{
V_{ub}V_{us}^* (a_1^{eff}X_1^{'} +a_2^{eff}X_2^{'} )\nonumber\\
&-& V_{tb}V_{ts}^* \left[ (a_4 +a_{10}+0.2(a_6 +a_8 ))X_1^{'}+
(2a_3 -2a_5 -\frac{1}{2}a_7 +\frac{1}{2}a_9 )X_{2u}^{'} \right.\nonumber\\
&+& \left. (a_3 +a_4 -a_5 +\frac{1}{2}a_7 -\frac{1}{2}a_9 -\frac{1}{2}a_{10}
+0.2(2a_6 -a_8 ))X_{2s}^{'}
\right] 
\},
\eea
\bea
{\cal M}_{iii}^{'}&=&\frac{G_F}{\sqrt{2}}
\left[
V_{cb}V_{cs}^* a_2^{eff}X_{2c}^{'}-V_{tb}V_{ts}^*
(a_3 -a_5 -a_7 +a_9 )X_{2c}^{'}
\right],
\eea
with 
\bea
X_1^{'}&=&\langle ~K^{*-} \mid (\bar s u)_{V-A} \mid 0\rangle ~
\langle ~\ete \mid (\bar u b)_{V-A} \mid B^- \rangle ~,\nonumber\\
X_{2q}^{'}&=&\langle ~\ete \mid (\bar q q)_{V-A} \mid 0\rangle ~
\langle ~K^{*-} \mid (\bar s b)_{V-A} \mid B^- \rangle ~.\nonumber
\eea
In the spirit of \cite{yangyd}, we can calculate ${\cal M}_{iv}^{\prime}$.
We take the wave function of $K^*$ as
\bea
\psi_{K^*} =\frac{1}{\sqrt{2}}\frac{I_C}{\sqrt{3}}
\phi_{K^*} (y)(\slash{\hskip -3mm}q+M_{K^*})~ 
e\!\!\!/,
\eea
where $\it e$ is polarization vector of $K^*$ and after a direct calculation,
we get
\bea
& &{\cal M}_{iv}^{\prime}=\frac{G_F}{\sqrt{2}}\alpha_s^2 C_{eff}C_F 32 
(V_{is}V_{ib}^*)\times\int dxdy\phi_{B}(x)\phi_{K^*}(y)\nonumber\\
& &F_1^i k_1^2
\{
\frac{M_B [(e\cdot k_1 )(q\cdot k_2 )-(e \cdot k_2 )(q\cdot k_1 )]
+M_{K^*}[(e\cdot k_2 )(p\cdot k_1 )-(e\cdot k_1 )(p\cdot k_2 )]}
{k_1\cdot k_2 k_1^2 k_2^2 }
\}\\
& &+F_2^i m_b 
\{
\frac{[(e\cdot k_1 )(p\cdot k_1 )(q\cdot k_2)
-(e\cdot k_2 )(p\cdot k_1 )(q\cdot k_1 )]+M_B M_{K^*}
[-(e \cdot k_1 )(k_1 \cdot k_2 )+(e \cdot k_2 )k_1^2 ]}
{k_1\cdot k_2 k_1^2 k_2^2 }
\}.\nonumber
\eea 
The numerical results are shown in Table 3, where we have taken 
$f_{K^*}=180 MeV$ and $\epsilon_B =0.06(0.07)$.
It is shown that all predictions on the decays are under
the upper limit of experimental data. Compared with the results of
\cite{chenghy1} and \cite{he} ,we have very different
predictions on $B^- \rightarrow \ete K^{*-} $ and 
$B^0 \rightarrow \ete K^{*0} $:
\bea
{\cal BR}(B^- \rightarrow \ete K^{*-}) = 3.18(2.61)\times 10^{-5},~~~~
{\cal BR}(B^0 \rightarrow \ete K^{*0}) = 3.33(2.74)\times 10^{-5}.
\eea
which is two orders larger than the predictions of 
\cite{chenghy1} and \cite{he}. If we use conventional amplitude used in the
literature \cite{chenghy1} and take into account the contribution of Mechanism
(iv) \cite{yangyd}, the results are listed in the sixth column of Table 3. We
find that the branching ratios
${\cal BR}(B^- \rightarrow \ete K^-)$ and ${\cal BR}(B^0 \rightarrow \ete K^0)$
all agree with the CLEO results. However, the predictions for
$B^- \rightarrow \ete K^{*-}$ and $B^0 \rightarrow \ete K^{*0}$ are
\bea
{\cal BR}(B^- \rightarrow \ete K^{*-}) = 7.41 \times 10^{-6},~~~~
{\cal BR}(B^0 \rightarrow \ete K^{*0}) = 7.29 \times 10^{-6},
\eea
where $\epsilon_B =0.06$, which are still one order larger than other
predictions \cite{chenghy1,he}. This character due to the mechanism (iv) can
be tested in the near future.

\section{Concluding Remarks}
In this paper, we first carefully examine the conventional decay mechanism
(using standard  effective weak Hamiltonian and BSW model) 
which contributes to the exclusive decay
$B^- \rightarrow \ete K^-$. Instead of using Dirac equation, we 
use an alternative PQCD method  to estimate
the hadronic matrix elements like
$\langle ~ \ete \mid \bar s \gamma_5 s \mid 0\rangle ~\langle ~K^- \mid \bar s b \mid B^- \rangle ~$
and get comparatively small results. We find that the standard
theorectical frame can only give 
$Br(B^- \rightarrow \ete K^-) \simeq 1\times 10^{-5}$
, which is too small to account for the experimental data. 

We also discussed some new mechanisms within the SM, we find that 
$f_{\ete}^c=140 MeV$
will predict a large branching ratio of $B^0 \rightarrow \ete K^{*0}$
,which is in disagreement with the experimental result, so the mechanism
through $b \rightarrow \bar c cs \rightarrow s \ete $
is unlikely to be dominant although its contributions to the decay
$B \rightarrow \ete K(K^*)$ are important. 
we combine the mechanisms of ref \cite{zhit}, ref \cite{yangyd}
and the conventional mechanism to interpret the experimental data. 
We find that the numerical
results of ${\cal BR}(B^{\pm} \rightarrow \ete K^{\pm})$ are in good agreement 
with the
experimental results and the predictions on the other decay channels are all 
under the experimental upper limit. It is interesting to note that we predict 
large branching fractions:
\bea
 {\cal BR}(B^- \rightarrow \ete K^{*-}) = 3.18(2.61)\times
10^{-5},\nonumber \\
{\cal BR}(B^0 \rightarrow \ete K^{*0}) = 3.33(2.74)\times
10^{-5}.\nonumber
\eea
which are still under the experimental upper limit but 
much larger than the predictions of \cite{chenghy1} and \cite{he}.
This can be tested in the near future.
From Table 3, we find that the large results  mainly come from
the two gluons fusion mechanism \cite{yangyd} which predict
the same order branching fraction for
$B \rightarrow \ete K$ and $B \rightarrow \ete K^* $
\\
{\large\bf Acknowledgment} \\
This work is supported in part by the National Natural Science Foundation of
China.

\newpage
\noindent
\begin{table}
\begin{tabular}{|c|c|c|c|c|}\hline
 $m_b =5.0 GeV$&$\epsilon_B =0.05$&$\epsilon_B =0.06$&
$\epsilon_B =0.07$&$\epsilon_B =0.08$\\ \hline
F& 0.31& 0.24& 0.20& 0.17 \\ \hline
Rm&0.190&0.192&0.193&0.195 \\ \hline
\end{tabular}
\vspace*{1cm}\\
\begin{tabular}{|c|c|c|c|c|}\hline
$m_b =4.9 GeV$&$\epsilon_B =0.05$&$\epsilon_B =0.06$&
$\epsilon_B =0.07$&$\epsilon_B =0.08$\\ \hline
F& 0.28& 0.22& 0.18& 0.16 \\ \hline
Rm&0.192&0.194&0.196&0.198 \\ \hline
\end{tabular}
\vspace*{1cm}\\
\begin{tabular}{|c|c|c|c|c|}\hline
$m_b =4.8 GeV$&$\epsilon_B =0.05$&$\epsilon_B =0.06$&
$\epsilon_B =0.07$&$\epsilon_B =0.08$\\ \hline
F& 0.26& 0.20& 0.17& 0.14 \\ \hline
Rm&0.194&0.197&0.199&0.201 \\ \hline
\end{tabular}

\caption{The PQCD estimations about the hadronic matrix element
$\langle ~K^- \mid \bar s \gamma_{\mu}b \mid B^- \rangle ~$
and the ratio of $\langle ~ \ete \mid \bar s \gamma_5 s \mid 0\rangle ~\langle ~K^- \mid \bar s b \mid B^- \rangle ~$
to $\langle ~ \ete \mid \bar s \gamma_{\mu} \gamma_5 s \mid 0\rangle ~\langle ~K^- \mid
\bar s \gamma^{\mu}b \mid B^- \rangle .$}
\end{table}

\vspace{2cm}

\newpage
\footnotesize
\begin{table}

\begin{tabular}{|c|c|c|c|c|c|}\hline
${\cal BR}(B^- \rightarrow \ete K^- )$&Mech.(iv) &Mech.(iv)+(conventional)&
\multicolumn{3}{|c|}{Mech.(iv)+(conventional)+Mech.(iii)}\\ 
\cline{4-6}
~&contribution&contribution&$f_{\ete}^c=10 MeV$& 
$f_{\ete}^c=-15 MeV$&$f_{\ete}^c=-30 MeV$\\ \hline
$\epsilon_B =0.05$&$2.65\times 10^{-5}$
&$4.67\times 10^{-5}$&$4.31\times 10^{-5}$
&$5.45\times 10^{-5}$&$6.51\times 10^{-5}$ \\ \hline

$\epsilon_B =0.06$&$2.0\times 10^{-5}$&$3.85\times 10^{-5}$
&$3.49\times 10^{-5}$&$4.63\times 10^{-5}$&$5.70\times 10^{-5}$ \\ \hline

$\epsilon_B =0.07$&$1.57\times 10^{-5}$&$3.32\times 10^{-5}$
&$2.96\times 10^{-5}$&$4.10\times 10^{-5}$&$5.17\times 10^{-5}$ \\ \hline

$\epsilon_B =0.08$&$1.30\times 10^{-5}$&$2.96\times 10^{-5}$
&$2.60\times 10^{-5}$&$3.74\times 10^{-5}$&$4.80\times 10^{-5}$ \\ \hline
\end{tabular}

\caption{Numerical results of ${\cal BR}(B^- \rightarrow \ete K^- )$
in different cases. For conventional mechanism(tree+penguin),
${\cal BR}(B^- \rightarrow \ete K^- )=0.9\times 10^{-5}$.}
\end{table}

\vspace{6cm}
\begin{table}
\vspace{6cm}

\begin{tabular}{|c|c|c|c|c|c|c|}\hline
Branching&Mech.(iv)&$\epsilon_B =0.06(0.07)$&
Ref[3]&Ref[10]&Ref[3]+&Exp.($10^{-5}$) [1]\\
Ratio&$\epsilon_B =0.06(0.07)$&$f_{\ete}^c=-15 MeV$& &
&Mech.(iv) & \\ \hline
$B^- \rightarrow \ete K^- $&$2.0(1.57)\times 10^{-5}$
&$4.63(4.10)\times10^{-5}$
&$5.69\times 10^{-5}$&$6.1\times 10^{-5}$
&$8.31\times 10^{-5}$&$6.5^{+1.5}_{-1.4}\pm 0.9 $ \\ \hline

$B^0 \rightarrow \ete K^0 $&$2.0(1.57)\times 10^{-5}$
&$4.03(3.66)\times 10^{-5}$&$5.19\times 10^{-5}$ 
&$5.69\times 10^{-5}$&$6.24\times 10^{-5}$ 
& $4.7^{+2.7}_{-2.0}\pm 0.9 $ \\ \hline

$B^- \rightarrow \ete K^{\star -}$&$1.52(1.12)\times 10^{-5}$
&$3.18(2.61)\times 10^{-5}$&$3.24\times 10^{-7}$
&$1.15\times 10^{-5}$&$7.41\times 10^{-6}$&$ <13$ \\ \hline

$B^0 \rightarrow \ete K^{\star 0}$&$1.52(1.12)\times 10^{-5}$
&$3.33(2.74)\times 10^{-5}$&$2.96\times 10^{-7}$ & 
&$7.29\times 10^{-6}$&$ <3.9$ \\ \hline

$B^- \rightarrow \eta K^-$&$3.46(2.77)\times 10^{-6}$
&$5.31(4.52)\times 10^{-6}$&$8.68\times 10^{-7}$
&$5.13\times 10^{-6}$&$5.29\times 10^{-6}$&$ <1.4$ \\ \hline

$B^0 \rightarrow \eta K^0$&$3.46(2.77)\times 10^{-6}$
&$5.02(3.90)\times 10^{-6}$&$3.28\times 10^{-7}$
&$4.72\times 10^{-6}$&$4.80\times 10^{-6}$&$ <3.3$ \\ \hline

$B^- \rightarrow \eta K^{\star -}$&$2.76(2.05)\times 10^{-6}$
&$3.44(4.20)\times 10^{-6}$&$3.70\times 10^{-6}$
&$1.31\times 10^{-5}$&$9.73\times 10^{-7}$&$ <3.0$ \\ \hline

$B^0 \rightarrow \eta K^{\star 0}$&$2.76(2.05)\times 10^{-6}$
&$2.40(3.09)\times 10^{-6}$&$2.44\times 10^{-6}$& 
&$4.21\times 10^{-7}$&$ <3.0$ \\ \hline
\end{tabular}
\caption{Numerical results for B decay channels of interest and compared with
 predictions of other references and experimental results.}
\end{table}

\begin{figure}[tb]
\vspace*{10cm}
\hspace*{-6cm}
\centerline{\epsfig{figure=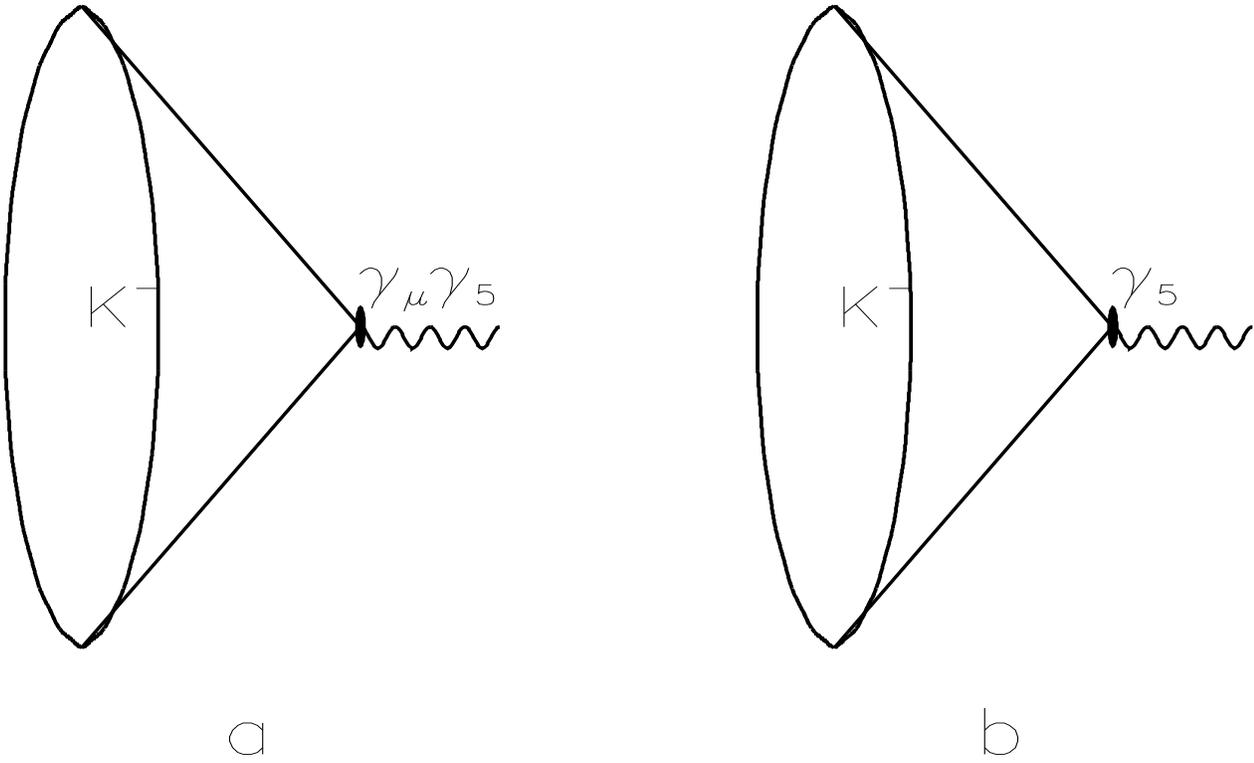,height=8cm,width=12cm,angle=0}}
\vspace*{-5cm}
\caption{\em Diagrams for the matrix elements $\langle K^- \mid \bar s
\gamma_{\mu} \gamma_5 u \mid 0 \rangle$(fig 1a) and
$\langle K^- \mid \bar s \gamma_5 u \mid 0 \rangle$(fig 1b).}
\end{figure}

\begin{figure}[tb]
\vspace*{11cm}
\hspace*{-6cm}
\centerline{\epsfig{figure=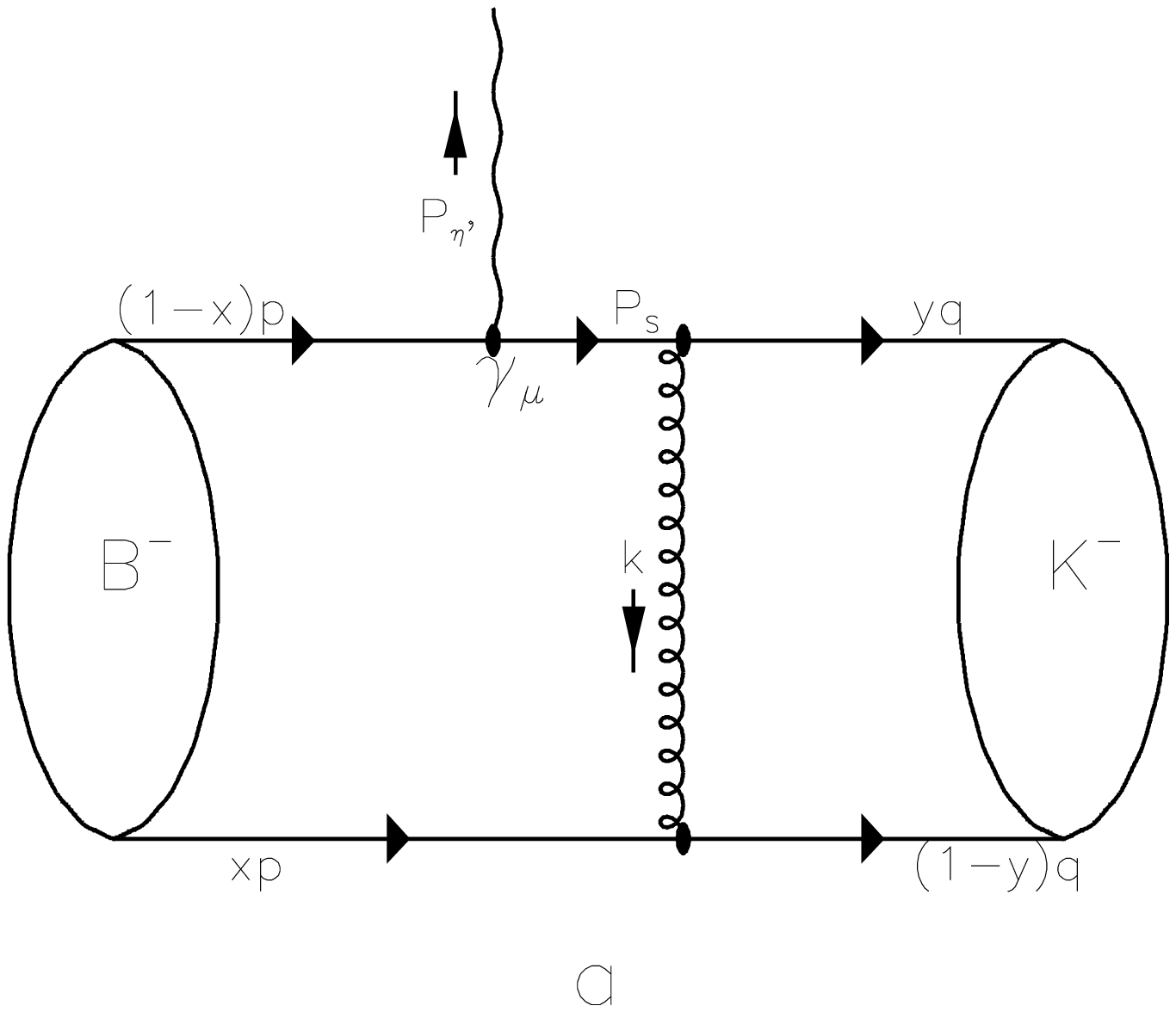,height=8cm,width=10cm,angle=0}}
\end{figure}

\begin{figure}[tb]
\vspace*{-12cm}
\hspace*{3cm}
\centerline{\epsfig{figure=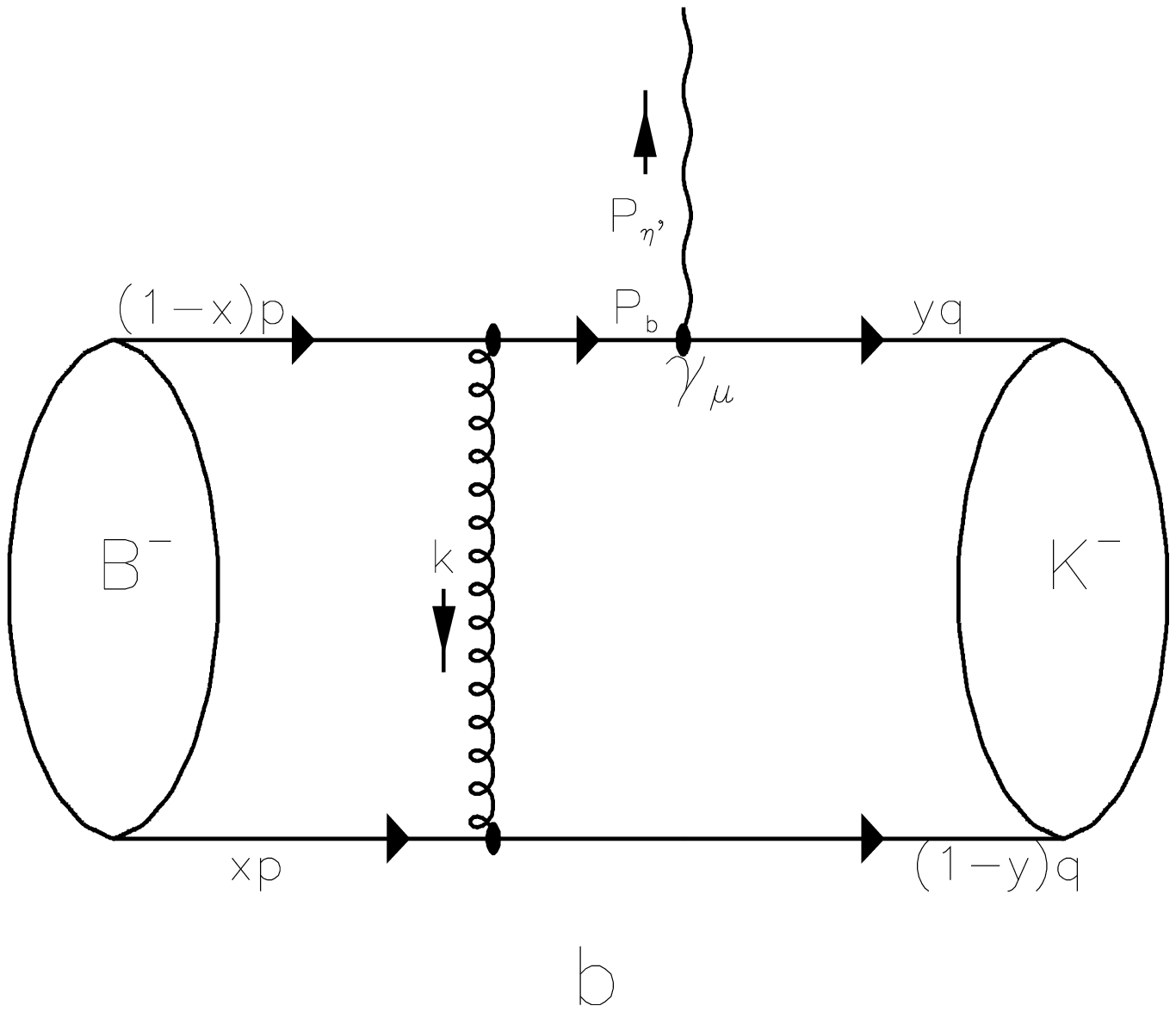,height=8cm,width=10cm,angle=0}}
\end{figure}

\begin{figure}[tb]
\vspace*{-4cm}
\caption{\em Leading twist diagrams for 
$\langle K^- \mid \bar s \gamma_{\mu}b \mid B^- \rangle$ in PQCD.}
\end{figure}

\newpage
~\\
\begin{figure}[tb]
\vspace*{4cm}
\hspace*{-6cm}
\centerline{\epsfig{figure=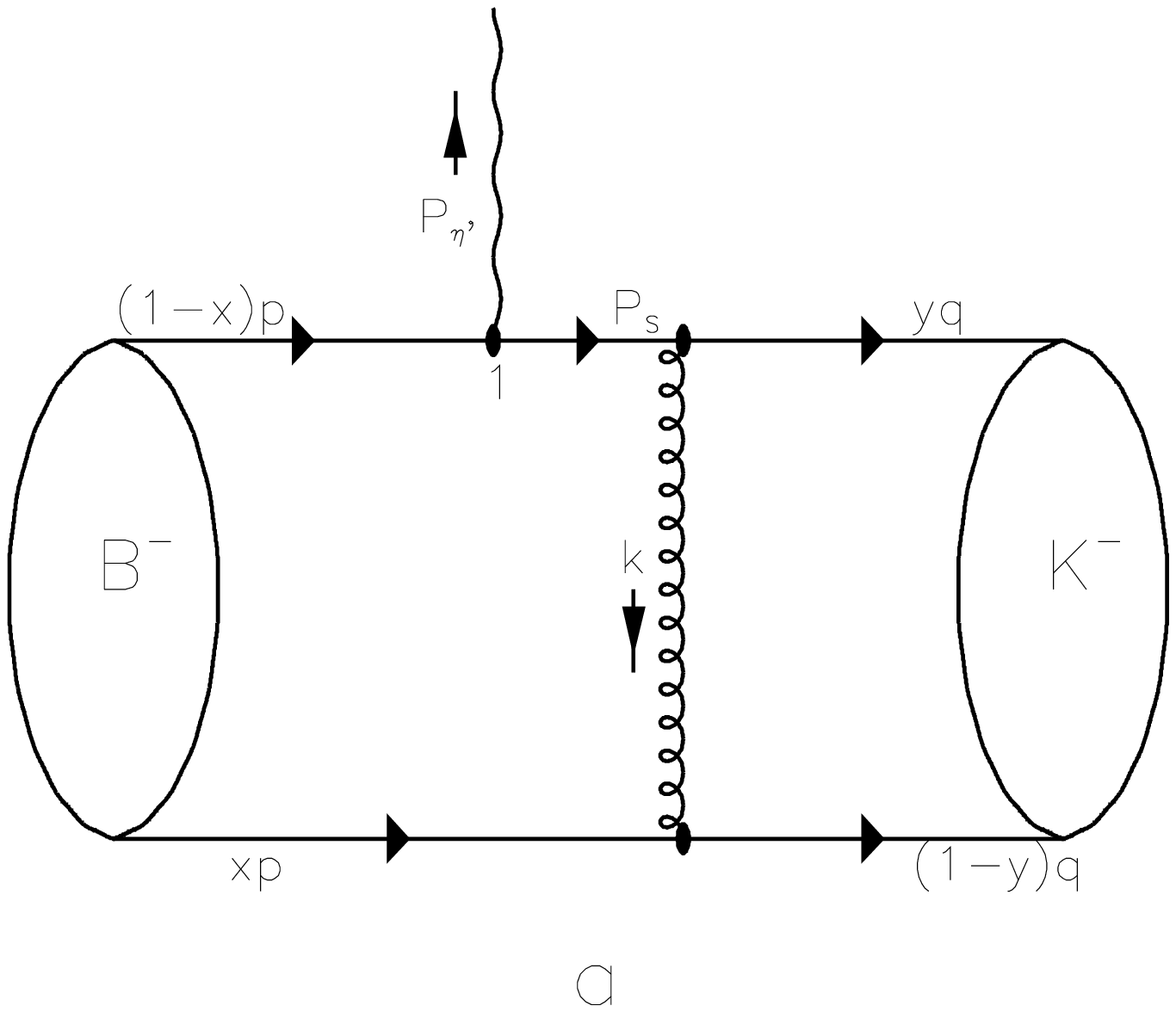,height=8cm,width=10cm,angle=0}}
\end{figure}

\begin{figure}[tb]
\vspace*{-13.5cm}
\hspace*{3cm}
\centerline{\epsfig{figure=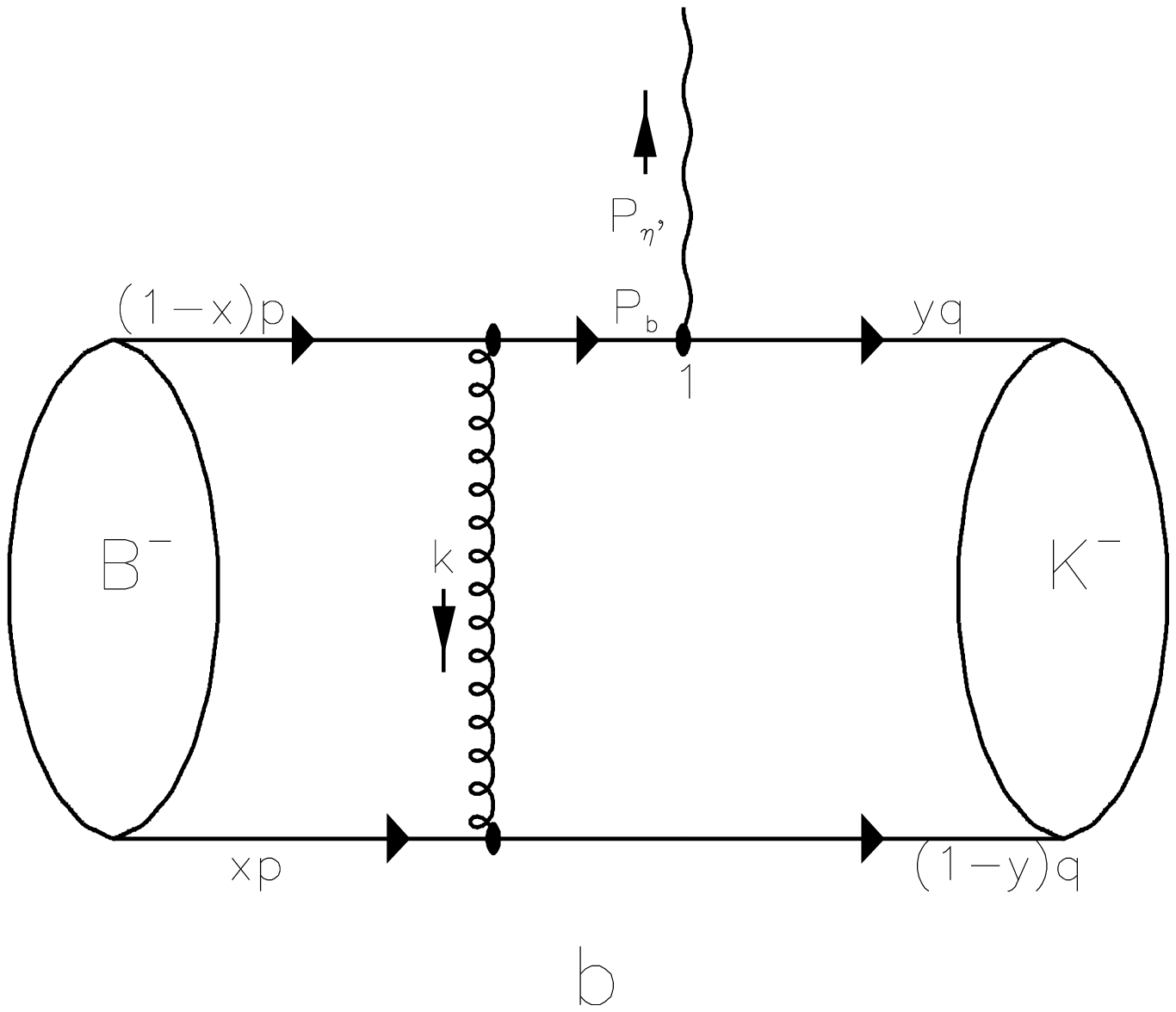,height=8cm,width=10cm,angle=0}}
\end{figure}

\begin{figure}[tb]
\vspace*{0cm}
\caption{\em Leading twist diagrams for 
$\langle K^- \mid \bar s b \mid B^- \rangle$ in PQCD.}
\end{figure}

\begin{figure}[tb]
\vspace*{10cm}
\hspace*{-5cm}
\centerline{\epsfig{figure=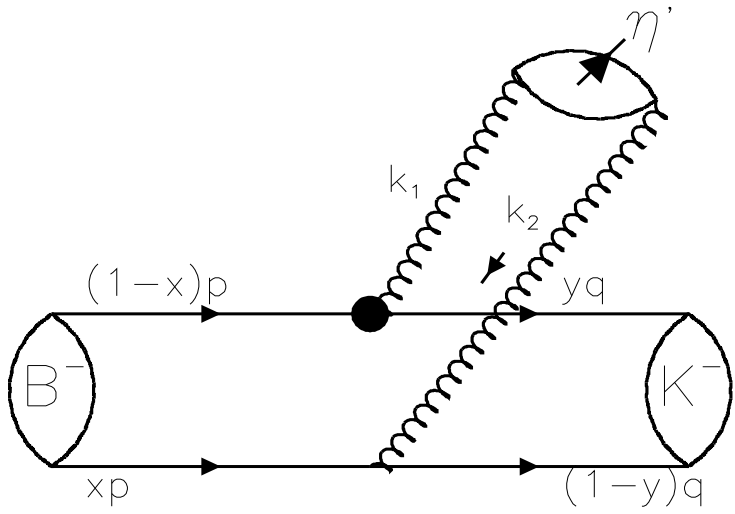,height=8cm,width=10cm,angle=0}}
\vspace*{-4cm}
\caption{\em Diagram for the two gluons fusion mechanism.} 
\end{figure}

\end{document}